\documentclass[preprintnumbers,amsmath,9pt,amssymb,floatfix,superscriptaddress,nofootinbib]{revtex4}
\usepackage{graphicx}
\usepackage{epsfig}
\usepackage{bm}
\usepackage{amsfonts}
\def\be {\begin{equation}}
\def\ee {\end{equation}}
\def\ba {\begin{eqnarray}}
\def\ea {\end{eqnarray}}

\begin{document}
\title{Power-Law and Logarithmic Entropy-Corrected Ricci Viscous Dark Energy and Dynamics of Scalar Fields}

 \author{\textbf{Antonio Pasqua}}\email{toto.pasqua@gmail.com} \affiliation{Department of Physics, University of Trieste, Via Valerio, 2 34127 Trieste, Italy}


\begin{abstract}
\textbf{Abstract:} In this work, I consider the logarithmic-corrected and the power-law corrected versions of the holographic dark energy (HDE) model in the
non-flat FRW universe filled with a viscous Dark Energy (DE) interacting with Dark Matter (DM).
I propose to replace the infra-red cut-off with the inverse of the Ricci scalar curvature $R$.
I obtain the equation of state (EoS) parameter $\omega_{\Lambda}$, the deceleration parameter $q$ and the evolution of energy density parameter $\Omega_D'$
 in the presence of interaction between DE and DM for both corrections. I study the correspondence of the logarithmic entropy corrected Ricci Dark Dnergy (LECRDE)
and power-law entropy corrected Ricci Dark Energy (PLECRDE) models with the the Modified Chaplygin Gas (MCG) and some scalar fields including tachyon, K-essence, dilaton and quintessence. I also make comparisons with previous results.\\ \\
\textbf{Keywords:} Dark energy; cosmology; scalar fields; cosmic acceleration.
\end{abstract}
\maketitle

\newpage

\section{INTRODUCTION}

Recent cosmological findings of Supernova Cosmology Project \cite{1,1-1}, Wilkinson Microwave Anisotropy Probe (WMAP)
\cite{2}, Sloan Digital Sky Survey (SDSS) \cite{3} and X-ray \cite{4,4-2} give convincing indication that the
observable universe is undergoing an accelerated expansion. To explain
this phenomenon the notion of dark energy (DE) with negative pressure was proposed.
 At present there are a lot of theoretical candidates of DE including tachyon, K-essence, dilaton, quintessence, H-essence
and DBI-essence, to name a few \cite{review,review2}. The simplest candidate for DE is the cosmological constant.
 From the point of view of quantum field theory, a cutoff at the Planck or electroweak scale leads to a cosmological
constant which is, respectively, $10^{123}$ or $10^{55}$ times larger than the observed value,
$\Lambda/8\pi G\sim10^{-47} GeV^4$. The absence of a fundamental symmetry which could set
the value of $\Lambda$ to either zero or a very small value leads to the cosmological constant
problem \cite{sahni}.

The complete and correct description of explanation of DE should come from the consistent theory of quantum gravity.
Such a theory does not yet exist and some approximations for this long-awaited theory are found including string theory and loop quantum gravity,
which are only effective theories. The string theory is based on some conjectures (like AdS/CFT) and the holographic principle,
according to the later the degrees of freedom of a physical system must scale according to the area and not by volume \cite{hooft}.
Using the holographic principle, a model of Holographic DE (HDE) was proposed \cite{li,li2}. Formally the idea of HDE goes like ``in quantum field theory
a short distance cut-off is related to a long distance cut-off due to the limit set by formation
of a black hole, namely, if $\rho_\Lambda$ is the quantum zero-point energy density caused by a short
distance cut-off, the total energy in a region of size $L$ should not exceed the mass of a black
hole of the same size, thus $L^3\rho_\Lambda\leq LM_p^2$. The largest $L$ allowed is the one saturating this
inequality, thus $\rho_\Lambda=3\alpha M_p^2L^{-2}$'' \cite{li}, where $\alpha$ is a constant and $M_p^2$ is the reduced Planck mass.

The power-law correction to entropy which appear in dealing with the entanglement of quantum fields in and out the horizon is given by \cite{das,das2,das3}:
\begin{equation}\label{1aa}
 S=c_0\Big( \frac{A}{a_1^2} \Big)[1+c_1f(A)],\ \ \ f(A)= \Big( \frac{A}{a_1^2} \Big)^{-\nu},
\end{equation}
where $c_0$ and $c_1$ are constants of order unity, $a_1$ is the
ultraviolet cut-off at the horizon and $\nu$ is a fractional
power which depends on the amount of mixing of ground and excited states. For large horizon area (i.e. $A\gg a_1^2$), the contribution of $f\left(A\right)$ is negligible and the mixed state entanglement
entropy asymptotically approaches the ground state (Bekenstein-Hawking) entropy \cite{das}.

Another useful form of entanglement entropy is \cite{power,power2}:
\begin{equation}\label{1a}
S_h=\frac{A_h}{4}\Big(1-K_\alpha A_h^{1-\frac{\alpha}{2}} \Big),\ \ \ K_\alpha=\frac{\alpha(4\pi)^{\frac{\alpha}{2}-1}}{(4-\alpha)r_c^{2-\alpha}},
\end{equation}
where $\alpha$ is a dimensionless constant and $A_h=4\pi R_h^2$ is the area of the horizon while $R_h$ is the radius of the horizon. Moreover, $r_c$ represents the cross-over scale. For entropy to be a well-defined quantity, I require that the condition $\alpha>0$ must be satisfied.

The quantum corrections provided to the entropy-area relationship leads to the curvature correction in the Einstein-Hilbert action and vice versa.
The logarithmic corrected entropy is  \cite{jamil5,jamil51,jamil52,jamil53}
\begin{equation}\label{1b}
S_h=\frac{A_h}{4}+\beta\log\Big( \frac{A_h}{4} \Big)+\gamma,
\end{equation}
where $\beta$ and $\gamma$ are constants. These
corrections arise in the black hole entropy in loop quantum gravity due to thermal equilibrium fluctuations and quantum fluctuations.

Based on the above quantum corrected entropies, the entropy corrected models of holographic models of DE were proposed recently.
Following Eq. (\ref{1a}), the energy density is given by \cite{power}:
\begin{equation}\label{1c}
\rho_\Lambda=3\alpha M_p^2 L^{-2}-\beta M_p^2 L^{-\gamma}.
\end{equation}
Similarly, following Eq. (\ref{1b}), the logarithmic corrected definition of energy density is \cite{wei,wei2,wei3,wei4}:
\begin{equation}\label{1d}
\rho_\Lambda=3\alpha M_p^2 L^{-2}+\gamma_1 L^{-4}\log (M_p^2 L^{2})+\gamma_2 L^{-4}.
\end{equation}

Gao et al. \cite{gao} pointed out that using the future event horizon as an infra-red cut-off for the HDE model leads to the causality problem.
They henceforth proposed the Ricci scalar curvature of FRW metric as a new cut-off $L=R^{-1/2}$ which can resolve not only the causality problem but also the coincidence problem. This model is called \textit{Ricci Dark Energy} (RDE). Gao et al. further pointed out that $\alpha\simeq0.46$
yields the correct DE density and equation of state today. Moreover, the RDE model is compatible with observational data from Supernovae, cosmic background radiation (CMB), baryon acoustic oscillations (BAO), gas mass fraction in galaxy
clusters, the history of the Hubble function, and the growth function \cite{pavon}. However, there is some criticism on RDE model:
 Kim et al. \cite{kim} pointed out that an accelerating phase of the RDE is that of a constant DE model. This implies
that the RDE may not be a new model of explaining the present accelerating
universe.

Motivated by the previous study of Gao et al. \cite{gao}, I consider the entropy corrected version of power-law entropy corrected Ricci DE as
\begin{eqnarray}
    \rho_{\Lambda}=3\alpha M_p^2R-\beta M_p^2R^{\gamma /2}, \label{5-1}
\end{eqnarray}
where $R$ represents the Ricci scalar curvature, which is given by:
\begin{eqnarray}
R=6\Big( \dot H+2H^2+\frac{k}{a^2} \Big),
\end{eqnarray}
where $H=\dot{a}\left(t\right)/a\left( t \right)$ is the Hubble parameter, $\dot{H}$ is the derivative of the Hubble parameter with respect to the cosmic time
 $t$, $a\left(t\right)$ is the scale factor (which is an arbitrary function of the cosmic time $t$ ) and $k$ is the curvature parameter (which has  dimension of $length^{-2}$ and describes the spatial geometry of spacetime). For $k=-1,\, 0,\, +1$ I obtain, respectively, an open, a flat or a closed FRW universe. \\
 Similarly, the energy density of logarithmic entropy corrected Ricci DE can be written as \cite{pasqua1}:
\begin{eqnarray}
    \rho_{\Lambda}=3\alpha M_p^2R+\gamma_1R^2 \log\Big(M_p^2/R\Big)+\gamma_2R^2.\label{5}
\end{eqnarray}
Ricci scalar curvature was considered as IR cut-off in different recent works \cite{r1,r2,r3,r4,r5,r6}.
Recently, the LECRDE and the PLECRDE ware proposed and their connections to various scalar field models of DE studied \cite{pasqua1,pasqua} without the effects of bulk viscosity.
In this paper, I study the bulk-viscosity in the both the DE models (\ref{5-1}) and (\ref{5}).

This paper is organized as follows. In Section II, I describe the physical contest I are working in and I derive the EoS parameter $\omega_{\Lambda}$,
the deceleration parameter $q$ and $\Omega_D'$ for our models in a non-flat universe. In Section III, I establish a correspondence between our model and the tachyon,
 K-essence, dilaton, quintessence and Modified Chaplygin Gas (MCG). The Conclusions of this work can be found in Section IV.

\section{INTERACTING MODEL IN A NON-FLAT UNIVERSE}

I assume the background spacetime to be Friedmann-Robertson-Walker (FRW) metric which is spatially homogeneous and isotropic and it has the line-element given by:
\begin{eqnarray}
    ds^2=-dt^2+a^2(t)\Big[\frac{dr^2}{1-kr^2} +r^2 (d\theta ^2 + \sin^2 \theta d\varphi ^2)\Big].\label{6}
\end{eqnarray}
where $t$ represents the cosmic time, $r$ is referred to the radial component and $\left(\theta, \phi \right)$ are the angular coordinates. \\
 The corresponding Friedmann equation takes the form:
\begin{eqnarray}
    H^2+\frac{k}{a^2}=\frac{1}{3M^2_p}( \rho _{\Lambda} + \rho _{m}),\label{7}
\end{eqnarray}
where $\rho _{\Lambda}$ and $\rho _{m}$ are the energy densities of DE and DM, respectively.

I also define the relative energy densities of matter, curvature and DE, respectively, as:
\begin{eqnarray}
    \Omega _m &=& \frac{\rho_m}{\rho _{cr}} = \frac{\rho _m}{3M_p^2H^2},\label{8}\\
    \Omega _k &=& \frac{\rho_k}{\rho _{cr}} = \frac{k}{H^2a^2},\label{9}\\
    \Omega _{\Lambda} &=& \frac{\rho_{\Lambda}}{\rho _{cr}} = \frac{\rho _{\Lambda}}{3M_p^2H^2},\label{10}
\end{eqnarray}
where $\rho_{cr} = 3M^2_pH^2$ is the critical density. $\Omega _k$ corresponds to the contribution of the spatial curvature to the total density.
Recent observations give a value of $\Omega _k \cong 0.02$ \cite{spergel}.

Using Eqs. (\ref{8}), (\ref{9}) and (\ref{10}) I can rewrite the Friedmann equation given in Eq. (\ref{7}) as:
\begin{eqnarray}
1 + \Omega _k = \Omega _{m} + \Omega _{\Lambda}.\label{11}
\end{eqnarray}
In order to preserve the Bianchi identity or local energy-momentum conservation law, i.e. $\nabla_{\mu}T^{\mu \nu}=0$,
the total energy density $\rho_{tot} = \rho_{\Lambda} + \rho_m$ must satisfy the following continuity equation:
\begin{eqnarray}
    \dot{\rho}+3H( 1+\omega )\rho=0,\label{12}
\end{eqnarray}
where:
\begin{eqnarray}
    \omega = \rho/p, \ \
    \rho=\rho_m + \rho_{\Lambda},\ \
    p=\bar{p}_{\Lambda}=p_{\Lambda}-3\xi H,
\end{eqnarray}
represents, respectively, the EoS parameter, the total energy density and the effective pressure of the DE. $\xi$ represents the bulk viscosity coefficient \cite{ren,ren2}.
I must remember that the DM is pressureless, which implies $p_m=0$. Following \cite{karami}, if I assume   $\xi = \varepsilon \rho_{\Lambda}H^{-1}$, where
$\varepsilon$ is a constant parameter, the total pressure $p$ can be written as:
\begin{eqnarray}
    p=(\omega_{\Lambda} -3\varepsilon )\rho_{\Lambda},
\end{eqnarray}
where $\omega_{\Lambda}= p_{tot}/\rho_{tot}$ represents the equation of state (EoS) parameter of the viscous DE.

By assuming an interaction between DM and DE, the two energy densities $\rho_{\Lambda}$ and $\rho_m$ for DE and DM are conserved separately and their conservation laws take the following form \cite{sheykhi,sheykhi2}:
\begin{eqnarray}
    \dot{\rho}_m+3H\rho_m &=& Q, \label{13}\\
        \dot{\rho}_{\Lambda}+3H\rho_{\Lambda}( 1 + \omega_{\Lambda})&=&-Q+9\xi H \rho_{\Lambda},  \label{14}
\end{eqnarray}
where $Q$ represents the interaction term which can be, in general, an arbitrary function of cosmological parameters, like the Hubble parameter $H$ and energy densities of DE and DM $\rho_{\Lambda}$ and $\rho_m$, i.e. $Q(H\rho_m,H\rho_{\Lambda})$. The simplest expression of $Q$ is \cite{umar,umar2,umar3,umar4,umar5,umar6}:
\begin{eqnarray}
    Q = 3b^2H(\rho _m + \rho _{\Lambda}).\label{15}
\end{eqnarray}
In literature, various forms of $Q$ have been proposed \cite{rashid,rashid2} and constraints on the coupling parameter $b^2$ have been investigated \cite{b2}.
Further high-resolution $N$-body simulations have shown that the structural properties of highly nonlinear cosmic structures, as e.g. their average concentration
at a given mass, could be significantly modified in the presence of an interaction between DE and DM \cite{salu}.
The strength of coupling parameter can, in fact, significantly modify the cosmic history by modifying the clustering properties of matter since the growth of DM density perturbations is much sensitive to the interaction \cite{maru,maru2}. The best way to motivate a suitable form of $Q$ should be from a consistent theory of quantum gravity or through a reconstruction scheme using the SNIa data \cite{sn,sn2}. However, so far, the interacting DE model is best-fitted with the observations \cite{wang}.

I now want to derive the expression for the EoS parameter $\omega_{\Lambda}$ for LECRDE and PLECRDE models.\\
Using the Friedamm equation given in Eq. (\ref{7}), the Ricci scalar curvature $R$ can be written as:
\begin{eqnarray}
    R=6\Big(  \dot{H} + H^2 + \frac{\rho_m+\rho_{\Lambda}}{3M_p^2} \Big). \label{17}
\end{eqnarray}
From the Friedmann equation given in Eq. (\ref{7}), I also derive:
\begin{eqnarray}
    \dot{H}&=&\frac{k}{a^2}-\frac{1}{2M_p^2}\Big( \rho + \bar{p}  \Big), \nonumber \\
     &=& \frac{k}{a^2} - \frac{1}{2M_p^2}\Big[ \rho_m + ( 1+\omega_{\Lambda}-3\varepsilon )\rho_{\Lambda} \Big].\label{dotH}
\end{eqnarray}
Adding Eqs. (\ref{7}) and (\ref{dotH}), I get:
\begin{eqnarray}    \dot{H}+H^2=\frac{1}{3M_p^2}\Big(\rho_m+\rho_{\Lambda}\Big)-\frac{\rho_m}{2M_p^2}-\frac{1}{2M_p^2}\Big( 1+\omega_{\Lambda} - 3\varepsilon  \Big)\rho_{\Lambda}.   \label{jamil20}
\end{eqnarray}
Therefore, the Ricci scalar $R$ given in Eq. (\ref{17}) can be rewritten as:
\begin{eqnarray}    R=\frac{\rho_m}{M_p^2}+\frac{3\rho_{\Lambda}}{M_p^2}\Big(\frac{1}{3}-\omega_{\Lambda}+3\varepsilon \Big). \label{jamil22}
\end{eqnarray}
I can now easily derive the expression of the EoS parameter $\omega_{\Lambda}$ from Eq. (\ref{jamil22}) :
\begin{eqnarray}    \omega_{\Lambda}=3\varepsilon-\frac{RM_p^2}{3\rho_{\Lambda}}+\frac{\Omega_{\Lambda}+\Omega_{m}}{3\Omega_{\Lambda}},\label{21}
\end{eqnarray}
where I used the fact that $\frac{\rho_{\Lambda}+\rho_{m}}{3\rho_{\Lambda}} = \frac{\Omega_{\Lambda}+\Omega_{m}}{3\Omega_{\Lambda}}$.

Substituting in Eq. (\ref{21}) the expression of the energy density $\rho_{\Lambda}$ given in Eq. (\ref{5}) and also using Eq. (\ref{11}), I obtain for the LECRDE model:
\begin{eqnarray}
    \omega_{\Lambda}=3\varepsilon -\frac{M_p^2/3}{3\alpha M_p^2+\gamma_1R\log\Big(M_p^2R^{-1}\Big)+\gamma_2R}+\frac{1}{3}\Big(\frac{1+\Omega_k}{\Omega_{\Lambda}}\Big).    \label{22}
\end{eqnarray}
Moreover, substituting Eq. (\ref{5-1}) in Eq. (\ref{21}) and using Eq. (\ref{11}), I obtain for the PLECRDE model:
\begin{eqnarray}
\omega_{\Lambda}=3\varepsilon -\frac{1}{3\Big(3\alpha -\beta R^{\gamma /2-1}\Big)}+\frac{1}{3}\Big(\frac{1+\Omega_k}{\Omega_{\Lambda}}\Big). \label{rplechde}
\end{eqnarray}
I now want to derive the expression for the evolutionary form of energy density parameter $\Omega_{\Lambda}$.\\
Using Eq. (\ref{14}), it is possible to obtain the following expression for the EoS parameter  $\omega_{\Lambda}$:
\begin{eqnarray}
    \omega_{\Lambda}= -1-\frac{\dot{\rho}_{\Lambda}}{3H\rho_{\Lambda}}-\frac{Q}{3H\rho_{\Lambda}}+9\varepsilon. \label{23}
\end{eqnarray}
Using the expression of the interaction term $Q$ given in Eq. (\ref{15}), the derivative of the DE energy density $\rho_{\Lambda}$ with respect to the cosmic time $t$ can be written as:
\begin{eqnarray}
    \dot{\rho}_{\Lambda}=3H\Big[ -\rho_{\Lambda}-\Big(\rho_m + \rho_{\Lambda}\Big)\Big(b^2+\frac{1}{3} \Big) + \frac{RM_p^2}{3} + 3\varepsilon \rho_{\Lambda} \Big].\label{24}
\end{eqnarray}
Dividing by the critical density $\rho_c=3H^2M_p^2$, Eq. (\ref{24}) becomes:
\begin{eqnarray}
    \frac{\dot{\rho}_{\Lambda}}{\rho_c}=3H\Big[ -\Omega_{\Lambda}-\Big(1 + \Omega_k\Big)\Big(b^2+\frac{1}{3} \Big) + \frac{R}{9H^2}+3\varepsilon\Omega_{\Lambda} \Big]=\dot{\Omega_{\Lambda}}+2\Omega_{\Lambda}\frac{\dot{H}}{H}.  \label{25}
\end{eqnarray}
Using Eq. (\ref{17}), I derive that the term $\frac{R}{9H^2}$ is equivalent to:
\begin{eqnarray}
    \frac{R}{9H^2}=\frac{2}{3}\Big( \frac{\dot{H}}{H^2} +2 + \Omega_k \Big).\label{26}
\end{eqnarray}
Substituting Eq. (\ref{26}) in Eq. (\ref{25}), it is possible to obtain the derivative of $\Omega_{\Lambda}$ with respect to the cosmic time $t$:
\begin{eqnarray}
\dot{\Omega}_{\Lambda}=2\frac{\dot{H}}{H}\Big(1-\Omega_{\Lambda}  \Big)+3H\Big[-\Omega_{\Lambda}-  \Big(1 + \Omega_k\Big)\Big(b^2-\frac{1}{3} \Big) + \frac{2}{3}+ 3\varepsilon\Omega_{\Lambda}\Big].\label{27}
\end{eqnarray}
Since   $\Omega_{\Lambda}'=\frac{d\Omega_{\Lambda}}{dx}= \frac{1}{H}\dot{\Omega}_{\Lambda}$ (where $x=\ln a$), Eq. (\ref{27}) becomes:
\begin{eqnarray}
H   \Omega_{\Lambda}'=2H'\Big(1-\Omega_{\Lambda}  \Big)+3H\Big[-\Omega_{\Lambda}-  \Big(1 + \Omega_k\Big)\Big(b^2-\frac{1}{3} \Big) + \frac{2}{3}+ 3\varepsilon\Omega_{\Lambda}\Big],\label{28}
\end{eqnarray}
which yields to:
\begin{eqnarray}
    \Omega_{\Lambda}'=\frac{2}{H}\Big(1-\Omega_{\Lambda}  \Big)+3\Big[-\Omega_{\Lambda}-  \Big(1 + \Omega_k\Big)\Big(b^2-\frac{1}{3} \Big) + \frac{2}{3}+3\varepsilon\Omega_{\Lambda} \Big].\label{29}
\end{eqnarray}
In Eq. (\ref{29}) I used the fact that:
\begin{eqnarray}
    H'=\frac{a'}{a}=1.\label{30}
\end{eqnarray}
For completeness, I also derive the deceleration parameter $q$, which is defined as:
\begin{eqnarray}
    q=-\frac{\ddot{a}a}{\dot{a}^2}=  -\frac{\ddot{a}}{aH^2}  =-1-\frac{\dot{H}}{H^2}. \label{33}
\end{eqnarray}
Taking the derivative respect to the cosmic time $t$ of the Friedmann equation given in Eq. (\ref{7}) and using Eqs. (\ref{11}), (\ref{13}) and (\ref{14}), the deceleration parameter $q$ given in Eq. (\ref{33}) can be rewritten as:
\begin{eqnarray}
    q=\frac{1}{2}\Big[1 + \Omega_k + 3\Omega_{\Lambda} \omega_{\Lambda}  \Big].\label{32}
\end{eqnarray}
Substituting the expression of the EoS parameter $\omega_{\Lambda}$ given in Eq. (\ref{22}), I obtain that the deceleration parameter $q$ for the LECRDE is given by:
\begin{eqnarray}
    q=1-\frac{1}{2}\left[ \frac{M_p^2\Omega _{\Lambda}}{3\alpha M_p^2+\gamma_1R \log\Big(M_p^2R^{-1}\Big)+\gamma_2R}\right]+ \Omega _k +\frac{9\varepsilon \Omega_{\Lambda}}{2}.\label{jamil26}
\end{eqnarray}
Instead, using Eq. (\ref{5-1}), I obtain that the deceleration parameter $q$ for the PLECRDE is given by:
\begin{eqnarray}
    q=1-\frac{1}{2}\left[\frac{\Omega _{\Lambda}}{3\alpha -\beta R^{\gamma /2-1}}\right]+ \Omega _k +\frac{9\varepsilon \Omega_{\Lambda}}{2}.
\end{eqnarray}
I can now derive some important quantities of the LECRDE and PLECRDE models in the limiting case for a flat dark dominated universe in HDE model, i.e. when $\gamma_1=\gamma_2=0$, $\Omega_{\Lambda}=1$ and $\Omega_k$=0 for the LECRDE and $\beta=0$, $\Omega_{\Lambda}=1$ and $\Omega_k$=0 for the PLECRDE.\\
The energy densities for DE given in Eqs. (\ref{5-1}) and (\ref{5}) both reduce to:
\begin{eqnarray}
\rho_{\Lambda}=3\alpha M_{p}^2 R.\label{34}
\end{eqnarray}
From the Friedmann equation given in Eq. (\ref{7}), I can derive the following relations for the Hubble parameter $H$ and the Ricci scalar curvature $R$, given,
respectively, by:
\begin{eqnarray}
H&=&\frac{6\alpha}{12\alpha-1}\Big(\frac{1}{t}\Big),\label{35} \\
R&=&\frac{36\alpha}{(12\alpha-1)^2}\Big(\frac{1}{t^2}\Big).\label{36}
\end{eqnarray}
Finally, the EoS parameter $\omega_{\Lambda}$ and the deceleration parameter $q$ reduce to:
\begin{eqnarray}
    \omega_{\Lambda}&=&\frac{1}{3}-\frac{1}{9\alpha}+3\varepsilon, \label{36-1}\\
    q&=&1-\frac{1}{6\alpha}+\frac{9\varepsilon}{2}. \label{36-2}
\end{eqnarray}
In the limiting case $\varepsilon =0$, Eqs. (\ref{36-1}) and (\ref{36-2}) reduce to:
\begin{eqnarray}
\omega_{\Lambda}&=&\frac{1}{3}-\frac{1}{9\alpha},\label{LEoS}\\
q&=&1-\frac{1}{6\alpha}.\label{Lq}
\end{eqnarray}
Using Eq. (\ref{36}) in Eq. (\ref{34}) I can write the energy density of DE as:
\begin{eqnarray}
\rho_{\Lambda}=3\alpha M_{p}^2
\left[\frac{36\alpha}{(12\alpha-1)^2} \cdot \frac{1}{t^2}\right].\label{rho}
\end{eqnarray}
From Eq. (\ref{LEoS}), I see that in the limiting case, the EoS
parameter of DE becomes a constant value in which for $\alpha<1/12$,
$\omega_{\Lambda}<-1$, where the phantom divide can be crossed.
Since the Ricci scalar diverges for $\alpha=1/12$, this value of
$\alpha$ can not be taken into account. From Eq. (\ref{Lq}), we
derive that the acceleration is started at $\alpha\leq 1/6$ where
the quintessence regime is also started ($\omega_{\Lambda} \leq
-1/3$).

\section{CORRESPONDENCE BETWEEN LECRDE AND PLECRDE MODELS AND SCALAR FIELDS}

I now establish a correspondence between the LECRDE and the PLECRDE models and the Modified Chaplygin Gas (MCG) and some scalar field as tachyon, K-essence, dilaton and quintessence.
This correspondence assumes a particular importantance because the scalar field models are an effective description of an underlying theory of DE.
Therefore, it is worthwhile to reconstruct the potential and the dynamics of scalar fields according the evolutionary form of the LECRDE and the PLECRDE models.
For this purpose, I first compare the energy density of the LECRDE and the PLECRDE models given in Eqs. (\ref{5-1}) and (\ref{5}) with the energy density of corresponding scalar
field model. Then, I equate the equations of state (EoS) parameter of scalar field models with the EoS parameter of the LECRDE and the PLECRDE models given in Eqs. (\ref{5-1}) and (\ref{22}).

\subsection{INTERACTING TACHYON MODEL}

The effective Lagrangian of the tachyon scalar field is motivated from open string field theory \cite{sen} and it is a successful candidate for cosmic acceleration.
It has the Lagrangian given by \cite{sen1}:
\begin{eqnarray}
L=-V(\phi)\sqrt{1-g^{\mu \nu}\partial _{\mu}\phi \partial_{\nu}\phi},\label{39}
\end{eqnarray}
where $V(\phi)$ is the potential of tachyon and $g^{\mu \nu}$ is the metric tensor. The energy density $\rho_{\phi}$ and pressure $p_{\phi}$ for the tachyon field are given, respectively, by:
\begin{eqnarray}
\rho_{\phi}&=&\frac{V(\phi)}{\sqrt{1-\dot{\phi}^2}},\label{40}\\
p_{\phi}&=& -V(\phi)\sqrt{1-\dot{\phi}^2}.\label{41}
\end{eqnarray}
Moreover, the EoS parameter $\omega_{\phi}$ of tachyon scalar field is given by:
\begin{eqnarray}
\omega_{\phi}=\frac{p_{\phi}}{\rho_{\phi}}=\dot{\phi}^2-1.\label{42}
\end{eqnarray}
In order to have a real value of the energy density $\rho_{\phi}$ for tachyon field, I must have that $-1 < \dot{\phi} < 1$. Consequently, from Eq. (\ref{42}), the EoS parameter $\omega_{\phi}$ of tachyon must lie in the range to $-1 < \omega_{\phi} < 0$. Hence, the tachyon field can interpret the accelerated expansion of the universe, but it can not enter the phantom regime, i.e. $\omega_{\Lambda}<-1$.\\
Comparing Eqs. (\ref{5}) and (\ref{40}), I derive the following expression for the potential $V( \phi )$ of the tachyon field:
\begin{equation}
    V(\phi)=\rho_{\Lambda} \sqrt{1-\dot{\phi}^2}.\label{43}
\end{equation}
Instead, equating Eqs. (\ref{22}) and (\ref{rplechde}) with Eq. (\ref{42}), I obtain the expressions of the kinetic energy term $\dot{\phi}^2$ for the LECRDE and the PLECRDE, respectively:
\begin{eqnarray}
\dot{\phi}^2&=& 1 + \omega_{\Lambda}=1+3\varepsilon-\frac{M_p^2/3}{3\alpha
M_p^2+\gamma_1R \log\Big(M_p^2/R\Big)+\gamma_2R} +
\frac{\Big(1+ \Omega _k  \Big)}{3\Omega _{\Lambda}},\label{44}\\
\dot{\phi}^2&=& 1 + \omega_{\Lambda}=1+3\varepsilon -\frac{1}{3\left(3\alpha -\beta R^{\gamma /2-1}\right)}+\frac{1}{3}\Big(\frac{1+\Omega_k}{\Omega_{\Lambda}}\Big).\label{44-1}
\end{eqnarray}
Making use of Eqs. (\ref{43}), (\ref{44}) and (\ref{44-1}), it is possible to write the potential of the tachyon for the LECRDE and the PLECRDE, respectively, as:
\begin{eqnarray}
    V( \phi  ) &=& \rho _{\Lambda} \sqrt{-3\varepsilon + \frac{M_p^2/3}{3\alpha M_p^2+\gamma_1R \log
    \Big(M_p^2/R\Big)+\gamma_2R} - \frac{\Big(1+ \Omega _k  \Big)}{3\Omega _{\Lambda}}},\label{45}\\
    V( \phi  ) &=&\rho _{\Lambda} \sqrt{-3\varepsilon +\frac{1}{3\Big(3\alpha-\beta R^{\gamma /2-1}\Big)}-\frac{1}{3}\Big(\frac{1+\Omega_k}{\Omega_{\Lambda}}\Big)}.\\\label{45-1}
\end{eqnarray}

I can derive from Eqs. (\ref{44}) and (\ref{45}) that the kinetic energy $\dot{\phi}^2$ and the potential $V( \phi )$ may exist if the following condition is satisfied:
\begin{equation}
    -1\leq \omega_{\Lambda} \leq 0.\label{46}
\end{equation}
Eq. (\ref{46}) implies that the phantom divide can not be crossed in a universe with accelerated expansion.\\
Using $\dot{\phi}=\phi'H$, Eqs. (\ref{44}) and (\ref{44-1}) become:
\begin{eqnarray}
 \phi'&=& \frac{1}{H}  \sqrt{1+3\varepsilon-\frac{M_p^2/3}{3\alpha M_p^2+\gamma_1R \log\Big(M_p^2/R\Big)+\gamma_2R} + \frac{\Big(1+ \Omega _k  \Big)}{3\Omega _{\Lambda}}},\label{47}\\
 \phi'&=& \frac{1}{H}  \sqrt{1+3\varepsilon -\frac{1}{3\Big(3\alpha -\beta R^{\gamma /2-1}\Big)}+\frac{1}{3}\Big(\frac{1+\Omega_k}{\Omega_{\Lambda}}\Big)}.\label{47-1}
\end{eqnarray}

The evolutionary forms of the tachyon scalar field for the LECRDE and PLECRDE can be obtained from Eqs. (\ref{47}) and (\ref{47-1}) as follow:
\begin{eqnarray}
    \phi(a) - \phi(a_0)&=&\int_{a_0}^a \frac{da}{aH}\sqrt{1+3\varepsilon-\frac{M_p^2/3}{3\alpha M_p^2+\gamma_1R \log\Big(M_p^2/R\Big)+\gamma_2R} + \frac{\Big(1+ \Omega _k  \Big)}{3\Omega _{\Lambda}}},\label{48}\\
    \phi(a) - \phi(a_0)&=&\int_{a_0}^a \frac{da}{aH}\sqrt{1+3\varepsilon -\frac{1}{3\Big(3\alpha -\beta R^{\gamma/2-1}\Big)}+\frac{1}{3}\Big(\frac{1+\Omega_k}{\Omega_{\Lambda}}\Big)},\label{48-1}
\end{eqnarray}
where $a_0$ represents the present day  value of the scale factor $a\left(t\right)$.\\
In order to solve the integrals given in Eqs. (\ref{48}) and (\ref{48-1}), I use the following relation:
\begin{equation}
\frac{da}{a H}=\frac{da}{a\cdot(\dot{a}/a)}=\frac{da}{\dot{a}}=\frac{da}{da/dt} = dt. \label{basic}
\end{equation}
In the limiting case for
flat dark dominated universe, i.e. when $\gamma_1=\gamma_2=0$,
$\Omega_{\Lambda} =1$ and $\Omega_k$=0 for the LECRDE model and
$\beta=0$, $\Omega_{\Lambda}=1$ and $\Omega_k$=0 for the PLECRDE
model, using Eq. (\ref{basic}), the scalar field reduces to:
\begin{equation}
\phi\left(t\right)=\sqrt{\frac{12\alpha -1+27\alpha \varepsilon}{9\alpha}}t,\label{49}
\end{equation}
and potential of the tachyon becomes:
\begin{equation}
    V\left( \phi  \right) =\rho _{\Lambda} \sqrt{-3\varepsilon +\frac{1}{9\alpha}-\frac{1}{3}}.\\\label{*1}
\end{equation}
Using the expression of energy density given in Eq. (\ref{rho}) in Eq.
(\ref{*1}), I get:
\begin{equation}
V(\phi)=\frac{108\alpha
^2M_p^2}{(12\alpha-1)^2}\sqrt{\frac{1-3\alpha-27\alpha
\varepsilon}{9\alpha}}\frac{1}{t^2},\label{50}
\end{equation}

In the limiting case of $\varepsilon = 0$, I obtain the following expressions of $\phi\left(t\right)$ and $V\left(\phi\right)$ in absence of bulk viscosity:
\begin{eqnarray}
\phi\left(t\right) &=& \sqrt{\frac{12\alpha -1}{9\alpha}}t,\label{a49} \\
V\left(\phi\right) &=& \frac{4M_p^2}{(12\alpha-1)}\sqrt{\alpha(1-3\alpha)}\frac{1}{\phi^2}.\label{50}
\end{eqnarray}
In this correspondence, the
scalar field exist provided that $\alpha
>1/12$, which shows that the phantom divide can not be achieved.

\subsection{INTERACTING K-ESSENCE MODEL}

Model of K-essence was proposed as a solution to the problem of small cosmological constant and late-time cosmic acceleration \cite{picon,picon2}. Its action is defined by
\begin{eqnarray}
S=\int d^4x \sqrt{-g}\,p(\phi, \chi ),\label{51}
\end{eqnarray}
where $p(\phi, \chi )$ corresponds to a pressure density and $g$ is the determinant of the metric tensor $g^{\mu \nu}$. According to the Lagrangian given in Eq. (\ref{51}), the pressure $p\left(\phi, \chi \right)$ and the energy density $\rho\left(\phi, \chi \right)$ of the field $\phi$ can be written, respectively, as:
\begin{eqnarray}
    p(\phi, \chi )&=&f(\phi)( -\chi+\chi ^2   ), \label{52}\\
        \rho(\phi, \chi )&=&f(\phi)(-\chi+3\chi ^2).\label{53}
\end{eqnarray}
where $f(\phi)$ is the potential of the K-essence model.\\
The EoS parameter $\omega_K$ of K-essence scalar field is given by:
\begin{eqnarray}
    \omega _K= \frac{p(\phi, \chi )}{\rho(\phi, \chi )}=\frac{\chi-1}{3\chi -1}.\label{54}
\end{eqnarray}
From Eq. (\ref{54}), I can see the phantom behavior of K-essence scalar field ($\omega_K < -1$) is obtained when the parameter $\chi$ lies in the range $1/3 < \chi < 1/2$.\\
In order to consider the K-essence field as a description of the interacting LECRDE and PLECRDE densities, I establish the correspondence between the K-essence EoS parameter, $\omega_K$, and the interacting LECRDE and PLECRDE EoS parameters.\\
The expressions of $\chi$ for the LECRDE and PLECRDE can be found equating  Eqs. (\ref{22}) and (\ref{rplechde}) with Eq. (\ref{54}), obtaining:
\begin{eqnarray}
    \chi &=& \frac{\omega_{\Lambda}-1}{3\omega_{\Lambda}-1}=\frac{-1+3\varepsilon-\frac{M_p^2/3}{3\alpha M_p^2+\gamma_1R \log(M_p^2/R)+\gamma_2R} + \frac{(1+ \Omega _k  )}{3\Omega _{\Lambda}}}{-1 +9\varepsilon - \frac{M_p^2}{3\alpha M_p^2+\gamma_1R \log(M_p^2/R)+\gamma_2R} + \frac{(1+ \Omega _k  )}{\Omega _{\Lambda}}},\label{55}\\
    \chi &=& \frac{\omega_{\Lambda}-1}{3\omega_{\Lambda}-1}=\frac{-1 + 3\varepsilon -\frac{1}{3(3\alpha -\beta R^{\gamma /2-1})} + \frac{1}{3}(\frac{1+\Omega_k}{\Omega_{\Lambda}})}{-1+9\varepsilon -\frac{1}{3\alpha -\beta R^{\gamma /2-1}}+(\frac{1+\Omega_k}{\Omega_{\Lambda}})}.\label{55-1}
\end{eqnarray}

Moreover, equating Eqs. (\ref{5}) and (\ref{53}), I obtain the following expression for $f\left(\phi \right)$:
\begin{eqnarray}
    f\left(\phi\right )=\frac{\rho_{\Lambda}}{\chi(3\chi-1)}.\label{56}
\end{eqnarray}
Using $\dot{\phi}^2=2\chi$ and $\dot{\phi}=\phi'H$ in Eqs. (\ref{55}) and (\ref{55-1}), I obtain:
\begin{eqnarray}
 \phi'&=& \frac{\sqrt{2}}{H}\sqrt{\frac{-1+3\varepsilon-\frac{M_p^2/3}{3\alpha M_p^2+\gamma_1R \log\Big(M_p^2/R\Big)+\gamma_2R} + \frac{\Big(1+ \Omega _k  \Big)}{3\Omega _{\Lambda}}}{-1+9\varepsilon- \frac{M_p^2}{3\alpha M_p^2+\gamma_1R \log\Big(M_p^2/R\Big)+\gamma_2R} + \frac{\Big(1+ \Omega _k  \Big)}{\Omega _{\Lambda}} }},\label{57}\\
 \phi'&=& \frac{\sqrt{2}}{H} \sqrt{\frac{-1+3\varepsilon -\frac{1}{3\Big(3\alpha -\beta R^{\gamma /2-1}\Big)} + \frac{1}{3}\Big(\frac{1+\Omega_k}{\Omega_{\Lambda}}\Big)}{-1 + 9\varepsilon -\frac{1}{3\alpha -\beta R^{\gamma /2-1}}+\Big(\frac{1+\Omega_k}{\Omega_{\Lambda}}\Big)}}.\label{57-1}
\end{eqnarray}
Integrating Eqs. (\ref{57}) and (\ref{57-1}), I find the evolutionary form of the K-essence scalar field:
\begin{eqnarray}
    \phi(a) -    \phi(a_0) &=& \sqrt{2} \int_{a_0}^a \frac{da}{aH}\sqrt{\frac{-1+3\varepsilon-\frac{M_p^2/3}{3\alpha M_p^2+\gamma_1R \log\Big(M_p^2/R\Big)+\gamma_2R} + \frac{\Big(1+ \Omega _k  \Big)}{3\Omega _{\Lambda}}}{-1+9\varepsilon- \frac{M_p^2}{3\alpha M_p^2+\gamma_1R \log\Big(M_p^2/R\Big)+\gamma_2R} + \frac{\Big(1+ \Omega _k  \Big)}{\Omega _{\Lambda}} }} ,\label{58}\\
    \phi(a) -    \phi(a_0) &=& \sqrt{2} \int_{a_0}^a \frac{da}{aH} \sqrt{\frac{3\varepsilon -\frac{1}{3\Big(3\alpha -\beta R^{\gamma /2-1}\Big)} + \frac{1}{3}\Big(\frac{1+\Omega_k}{\Omega_{\Lambda}}\Big)-1}{9\varepsilon -\frac{1}{3\alpha -\beta R^{\gamma /2-1}}+\Big(\frac{1+\Omega_k}{\Omega_{\Lambda}}\Big)-1}} .\label{58-1}
\end{eqnarray}
In the limiting case of a flat dark dominated universe, i.e. when
$\gamma_1=\gamma_2=0$, $\Omega_{\Lambda}=1$ and $\Omega_k$=0 for the
LECRDE model and $\beta=0$, $\Omega_{\Lambda}=1$ and $\Omega_k$=0
for the PLECRDE model, and using Eq. (\ref{basic}), the scalar
field of K-essence reduces to:
\begin{eqnarray}
\phi(t)=\sqrt{\frac{2\Big(-6\alpha-1+27\varepsilon \alpha\Big)}{3\Big(27\varepsilon \alpha -1\Big)}}t.\label{fi1}
\end{eqnarray}
Moreover, Eqs. (\ref{55}) and (\ref{55-1}), in the limiting cases, become:
\begin{eqnarray}
    \chi =\frac{27\varepsilon \alpha -1-6\alpha}{3(27\varepsilon \alpha-1)}.\label{55-11}
\end{eqnarray}
Using Eq. (\ref{55-11}) along with Eq. (\ref{rho}) in Eq. (\ref{56}),  I get the potential term as:
\begin{eqnarray}
f(\phi)=\frac{54\alpha M_p^2}{\Big(12\alpha-1\Big)^2}\frac{\Big(1-27\varepsilon \alpha\Big)^2}{6\alpha+1-27\varepsilon \alpha}\frac{1}{t^2}.\label{fi2}
\end{eqnarray}
In the limiting case of $\varepsilon=0$, I obtain that the expressions of $\phi\left(t\right)$ and $f\left(\phi\right)$ reduce, respectively, to:
\begin{eqnarray}
\phi\left(t\right)&=&\sqrt{\frac{12\alpha+2}{3}}t,\label{59}\\
f\left(\phi\right)&=&\frac{36\alpha M_p^2}{(12\alpha-1)^2}\frac{1}{\phi^2}.\label{60}
\end{eqnarray}
I see that our universe may behave in all accelerated regimes
(phantom and quintessence), since $\alpha$ can assume all the
possible values.

\subsection{INTERACTING DILATON MODEL}

Dilaton model arises as a low-energy limit of string theory and is found to be a useful candidate of DE \cite{fooqi}. The expressions of its pressure and energy density are, respectively:
\begin{eqnarray}
p_D&=&-\chi +ce ^{\lambda \phi}\chi^2, \label{61}  \\
\rho_D&=&-\chi +3ce ^{\lambda \phi}\chi^2,\label{62}
\end{eqnarray}
where $c$ and $\lambda$ are two positive constants and $2\chi=\dot{\phi}^2$. The negative coefficient of the kinematic term of the dilaton field in Einstein frame produces a phantom-like behavior for dilaton field. \\
The EoS parameter $\omega_D$ for the dilaton scalar field is given by:
\begin{eqnarray}
    \omega _D= \frac{p_D}{\rho_D}=\frac{-1 +c e ^{\lambda \phi}\chi}{-1 +3c e ^{\lambda \phi}\chi}.\label{63}
\end{eqnarray}
In order to consider the dilaton field as a description of the interacting LECRDE and PLECRDE densities, I establish the correspondence between the dilaton EoS parameter, $\omega_D$, and the EoS parameter $\omega_{\Lambda}$ of the LECRDE and PLECRDE models. By equating Eqs. (\ref{22}) and (\ref{rplechde}) with Eq. (\ref{63}), I obtain:
\begin{eqnarray}
    ce ^{\lambda \phi}\chi &=& \frac{\omega_{\Lambda}-1}{3\omega_{\Lambda}-1}=\frac{-1+3\varepsilon-\frac{M_p^2/3}{3\alpha M_p^2+\gamma_1R \log\Big(M_p^2/R\Big)+\gamma_2R} + \frac{\Big(1+ \Omega _k  \Big)}{3\Omega _{\Lambda}}}{-1+9\varepsilon- \frac{M_p^2}{3\alpha M_p^2+\gamma_1R \log\Big(M_p^2/R\Big)+\gamma_2R} + \frac{\Big(1+ \Omega _k  \Big)}{\Omega _{\Lambda}}}, \label{64}\\
    ce ^{\lambda \phi}\chi &=& \frac{\omega_{\Lambda}-1}{3\omega_{\Lambda}-1}=\frac{-1+3\varepsilon -\frac{1}{3\Big(3\alpha -\beta R^{\gamma /2-1}\Big)} + \frac{1}{3}\Big(\frac{1+\Omega_k}{\Omega_{\Lambda}}\Big)}{-1+9\varepsilon -\frac{1}{3\alpha -\beta R^{\gamma /2-1}}+\Big(\frac{1+\Omega_k}{\Omega_{\Lambda}}\Big)}. \label{64-1}
\end{eqnarray}

Using the relation $\dot{\phi}^2=2\chi$, I can rewrite Eqs. (\ref{64}) and (\ref{64-1}) as:
\begin{eqnarray}
    e^{\lambda \phi/2} \dot{\phi}&=& \sqrt{ \frac{-2+6\varepsilon-\frac{2M_p^2/3}{3\alpha M_p^2+\gamma_1R \log\Big(M_p^2/R\Big)+\gamma_2R} + \frac{2\Big(1+ \Omega _k  \Big)}{3\Omega _{\Lambda}}}   {c\Big(-1+9\varepsilon- \frac{M_p^2}{3\alpha M_p^2+\gamma_1R \log\Big(M_p^2/R\Big)+\gamma_2R} + \frac{\Big(1+ \Omega _k  \Big)}{\Omega _{\Lambda}}\Big) }    },\label{65}\\
    e^{\lambda \phi/2} \dot{\phi}&=& \sqrt{  \frac{-2+6\varepsilon -\frac{2}{9\alpha-3\beta R^{\gamma /2-1}} + \frac{2}{3}\Big(\frac{1+\Omega_k}{\Omega_{\Lambda}}\Big)}{c\Big(-1+9\varepsilon -\frac{1}{3\alpha -\beta R^{\gamma /2-1}}+\Big(\frac{1+\Omega_k}{\Omega_{\Lambda}}\Big)\Big)}     }.\label{65-1}
\end{eqnarray}

Integrating Eq. (\ref{65}) with respect to the scale factor $a(t)$, I obtain:
\begin{eqnarray}
    e^{\frac{\lambda \phi(a)}{2}} &=& e^{\frac{\lambda \phi(a_0)}{2}}+  \frac{\lambda}{2\sqrt{c}}\int_{a_0}^a\frac{da}{aH}\sqrt{  \frac{-2+6\varepsilon-\frac{2M_p^2/3}{3\alpha M_p^2+\gamma_1R \log\Big(M_p^2/R\Big)+\gamma_2R} + \frac{2\Big(1+ \Omega _k  \Big)}{3\Omega _{\Lambda}}}   {-1+9\varepsilon- \frac{M_p^2}{3\alpha M_p^2+\gamma_1R \log\Big(M_p^2/R\Big)+\gamma_2R} + \frac{\Big(1+ \Omega _k  \Big)}{\Omega _{\Lambda}}}},\label{66}\\
    e^{\frac{\lambda \phi(a)}{2}} &=& e^{\frac{\lambda \phi(a_0)}{2}}+\frac{\lambda}{2\sqrt{c}}\int_{a_0}^a\frac{da}{aH}  \sqrt{  \frac{-2+6\varepsilon -\frac{2}{3\Big(3\alpha -\beta R^{\gamma /2-1}\Big)} + \frac{2}{3}\Big(\frac{1+\Omega_k}{\Omega_{\Lambda}}\Big)}{c\Big(-1+9\varepsilon -\frac{1}{3\alpha -\beta R^{\gamma /2-1}}+\Big(\frac{1+\Omega_k}{\Omega_{\Lambda}}\Big)\Big)}     }.\label{66-1}
\end{eqnarray}

The evolutionary form of the dilaton scalar field is then given by:
\begin{eqnarray}
    \phi\left(a\right) &=&  \frac{2}{\lambda}\ln \Big[e^{\frac{\lambda \phi(a_0)}{2}}
   +   \frac{\lambda}{\sqrt{2c}}  \int_{a_0}^a \frac{da}{aH}\sqrt{  \frac{-1+3\varepsilon-\frac{M_p^2/3}{3\alpha M_p^2+\gamma_1R \log\Big(M_p^2/R\Big)+\gamma_2R} + \frac{\Big(1+ \Omega _k  \Big)}{3\Omega _{\Lambda}}}   {-1+9\varepsilon- \frac{M_p^2}{3\alpha M_p^2+\gamma_1R \log\Big(M_p^2/R\Big)+\gamma_2R} + \frac{\Big(1+ \Omega _k  \Big)}{\Omega _{\Lambda}}}}\Big],\label{67}\\
    \phi\left(a\right) &=&  \frac{2}{\lambda}\ln \Big[e^{\frac{\lambda \phi(a_0)}{2}}+ \frac{\lambda}{\sqrt{2c}}  \int_{a_0}^a \frac{da}{aH}\sqrt{  \frac{-1+3\varepsilon -\frac{1}{3\Big(3\alpha -\beta R^{\gamma /2-1}\Big)} + \frac{1}{3}\Big(\frac{1+\Omega_k}{\Omega_{\Lambda}}\Big)}{-1+9\varepsilon -\frac{1}{3\alpha -\beta R^{\gamma /2-1}}+\Big(\frac{1+\Omega_k}{\Omega_{\Lambda}}\Big)} }\Big].\label{67-1}
\end{eqnarray}

In the limiting case of $\gamma_1=\gamma_2=0$, $\Omega_{\Lambda}=1$
and $\Omega_k$=0 for the LECRDE model and $\beta=0$,
$\Omega_{\Lambda}=1$ and $\Omega_k$=0 for the PLECRDE model, in a
flat dark dominated universe, and using Eq. (\ref{basic}) the scalar
field of dilaton field reduces to the following form:
\begin{eqnarray}
\phi(t)=\frac{2}{\lambda}\ln{\left[ \lambda t\sqrt{\frac{6\alpha+1-27\varepsilon \alpha}{6c(1-27\varepsilon \alpha)}}\right]}.\label{68-1}
\end{eqnarray}
In the limiting case of $\varepsilon = 0$, I obtain the following result:
\begin{eqnarray}
\phi(t)=\frac{2}{\lambda}\ln{\left[\lambda t\sqrt{\frac{1+6\alpha}{6c}}\right]}.\label{68}
\end{eqnarray}
I see that all values of $\alpha$ are permitted and, therefore, by this correspondence, the universe may behave in phantom and quintessence regime.

\subsection{QUINTESSENCE}

Quintessence is described by a time-dependent and homogeneous scalar field $\phi$ minimally coupled to gravity which has a potential $V(\phi)$ that leads to the accelerating universe. The action for quintessence is given by \cite{cope}:
\begin{eqnarray}
    S=\int d^4x \sqrt{-g}\,\Big[-\frac{1}{2}g^{\mu \nu} \partial _{\mu} \phi   \partial _{\nu} \phi - V( \phi )  \Big],\label{69}
\end{eqnarray}
where $g$ represents the determinant of $g^{\mu \nu}$.\\
The energy-momentum tensor $T_{\mu \nu}$ of the field is derived by varying the action $S$ given in Eq. (\ref{69}) with respect to the metric tensor $g^{\mu \nu}$:
\begin{eqnarray}
T_{\mu \nu}=\frac{2}{\sqrt{-g}} \frac{\delta S}{\delta g^{\mu \nu}},\label{70}
\end{eqnarray}
which yields to:
\begin{eqnarray}
    T_{\mu \nu}=\partial _{\mu} \phi   \partial _{\nu} \phi - g_{\mu \nu}\Big[\frac{1}{2}g^{\alpha \beta} \partial _{\alpha} \phi   \partial _{\beta} \phi + V( \phi )  \Big].\label{71}
\end{eqnarray}
In a FRW background, the energy density and pressure of the quintessence scalar field $\phi$ are given, respectively, by:
\begin{eqnarray}
    \rho_Q&=&-T_0^0=\frac{1}{2}\dot{\phi}^2+V(\phi),\label{72}\\
    p_Q&=&T_i^i=\frac{1}{2}\dot{\phi}^2-V(\phi).\label{73}
\end{eqnarray}
The EoS parameter $\omega_Q$ for the quintessence scalar field is given by:
\begin{eqnarray}    \omega_Q=\frac{p_Q}{\rho_Q}=\frac{\dot{\phi}^2-2V(\phi)}{\dot{\phi}^2+2V(\phi)}.\label{74}
\end{eqnarray}
I find from Eq. (\ref{74}) that, when $\omega_Q < -1/3$, the universe accelerates for $\dot{\phi}^2<V(\phi)$. Hence the scalar potential must be shallow enough in order the field can evolve slowly along the potential.\\
The variation with respect to $\phi$ of the quintessence action given in Eq. (\ref{69}) yields to:
\begin{eqnarray}
    \ddot{\phi} + 3H\dot{\phi}+V_{,\phi} = 0.
\end{eqnarray}
I now establish the correspondence between the interacting scenario and the quintessence DE model: equating Eq. (\ref{74}) with the EoS parameter given in Eq. (\ref{22}), i.e.  $\omega_Q=\omega_{\Lambda}$, and equating Eqs. (\ref{72}) and (\ref{5}), i.e.  $\rho_Q=\rho_{\Lambda}$, I obtain:
\begin{eqnarray}
    \dot{\phi}^2&=&(1+\omega_{\Lambda})\rho_{\Lambda}, \label{75}\\
    V( \phi ) &=& \frac{1}{2}(1-\omega_{\Lambda})\rho_{\Lambda}.\label{76}
\end{eqnarray}
Substituting Eqs. (\ref{22}) and (\ref{rplechde}) into Eqs. (\ref{75}) and (\ref{76}), the kinetic energy term $\dot{\phi}^2$ and the quintessence potential $V\left( \phi \right)$ for the LECRDE and PLECRDE can be easily found as follow:
\begin{eqnarray}
    \dot{\phi}^2&=&\rho_{\Lambda}\Big[1+3\varepsilon-\frac{M_p^2/3}{3\alpha M_p^2+\gamma_1R \log\Big(M_p^2/R\Big)+\gamma_2R} + \frac{(1+ \Omega _k )}{3\Omega _{\Lambda}}\Big],\label{77}\\
    V( \phi ) &=&\frac{\rho_{\Lambda}}{2}  \Big[ 1-3\varepsilon+\frac{M_p^2/3}{3\alpha M_p^2+\gamma_1R \log\Big(M_p^2/R\Big)+\gamma_2R} - \frac{(1+ \Omega _k  )}{3\Omega _{\Lambda}}  \Big],\label{78}\\
    \dot{\phi}^2&=&\rho_{\Lambda}\Big[1+3\varepsilon -\frac{1}{3\Big(3\alpha-\beta R^{\gamma /2-1}\Big)}+\frac{1}{3}\Big(\frac{1+\Omega_k}{\Omega_{\Lambda}}\Big)\Big],\label{eqn78+1}\\
    V( \phi) &=&\frac{\rho_{\Lambda}}{2}  \Big[ 1  -  3\varepsilon +\frac{1}{3\Big(3\alpha-\beta R^{\gamma /2-1}\Big)}-\frac{1}{3}\Big(\frac{1+\Omega_k}{\Omega_{\Lambda}}\Big)  \Big].\label{eqn78+2}
\end{eqnarray}
Integrating (\ref{77}) and (\ref{eqn78+1}), and using the relation $\dot{\phi}=\phi' H$ and Eq. (\ref{rho}), it is possible to obtain the evolutionary form of the quintessence scalar fields as:
\begin{eqnarray}
\phi(a) - \phi(a_0)&=& \int_{a_0}^{a}\frac{da}{a H}\sqrt{\rho_{\Lambda}\left(1+3\varepsilon-\frac{M_p^2/3}{3\alpha M_p^2+\gamma_1R \log\Big(M_p^2/R\Big)+\gamma_2R} + \frac{\Big(1+ \Omega _k
\Big)}{3\Omega _{\Lambda}}\right)},\label{79} \\
\phi(a) - \phi(a_0)&=& \int_{a_0}^{a}\frac{da}{a H}\sqrt{\rho_{\Lambda}\left(1+3\varepsilon -\frac{1}{3\Big(3\alpha-\beta R^{\gamma /2-1}\Big)}+\frac{1}{3}\Big(\frac{1+\Omega_k}{\Omega_{\Lambda}}\Big)\right)},\label{79-1}
\end{eqnarray}
In the limiting case of $\gamma_1=\gamma_2=0$, $\Omega_{\Lambda}=1$
and $\Omega_k$=0 for the LECRDE model and $\beta=0$,
$\Omega_{\Lambda}=1$ and $\Omega_k$=0 for the PLECRDE model, in a
flat dark dominated universe, and using Eq. (\ref{basic}) the scalar
field of quintessence reduce to:
\begin{eqnarray}
\phi\left(t\right) - \phi\left(t_0\right)= \int_{t_0}^{t}\frac{dt}{t}\frac{6\alpha
M_p}{\Big(12\alpha -1\Big)}\sqrt{\frac{12\alpha-1+27\alpha
\varepsilon}{3\alpha}},\label{79-11}
\end{eqnarray}
where I used Eq. (\ref{rho}). \\
Eq. (\ref{79-11}) yields:
\begin{eqnarray}
\phi\left(t\right) = \frac{6\alpha M_p}{\Big(12\alpha -1\Big)}\sqrt{\frac{12\alpha-1+27\alpha \varepsilon}{3\alpha}} \ln \left( t \right). \label{80-1}
\end{eqnarray}
Moreover, the potential becomes:
\begin{eqnarray}
V(\phi) = \frac{54\alpha^2 M_p^2}{\Big( 12\alpha - 1 \Big)^2} \Big( \frac{6\alpha - 27\alpha \varepsilon +1}{9\alpha} \Big)\frac{1}{t^2}. \label{81-1}
\end{eqnarray}

In the limiting case $\varepsilon = 0$, Eqs. (\ref{80-1}) and Eqs.
(\ref{81-1}) reduce to:
\begin{eqnarray}
\phi\left(t\right) &=& \frac{6\alpha M_p}{\sqrt{3\alpha(12\alpha-1)}}\ln{(t)},\label{80}\\
V\left(\phi\right) &=& \frac{6\alpha(6\alpha+1)}{(12\alpha-1)^2}M_p^2\exp{\Big[\frac{-\sqrt{3\alpha(12\alpha-1)}}{3\alpha
M_p}\phi\Big]}.\label{81}
\end{eqnarray}

The potential exists for all values of $\alpha >1/12$ (quitessence regime). The potential has also been obtained by power-law expansion of scale factor.

\subsection{MODIFIED CHAPLYGIN GAS}
In this Section, I describe the correspondence between the Modified Chaplygin Gas (MCG) and our models.\\
One of the suggested candidates for DE is the GCG, which represents the generalization of the Chaplygin gas \cite{Kamenshchik}. GCG has the favourable property of interpolating the evolution of the universe from the dust to the accelerated phase, hence it fits better the observational data \cite{li2010}. The GCG and its further generalization have been widely studied in the literature \cite{thakur, gonzales, chatto, benaoum, li2009}.\\
The GCG is defined as \cite{bento, bento-2}:
\begin{eqnarray}
	p_{\Lambda}=-\frac{D}{\rho_{\Lambda}^{\theta}}, \label{gcg1}
\end{eqnarray}
where $D$ and $\theta$ are constants (with $D$ also positive defined). The Chaplygin gas is obtained in the limiting case $\theta = 1$. Gorini et al. \cite{gorini} showed that the matter power spectrum is compatible with the observed one only when $\theta < 10^{-5}$, which means that the GCG is practically indistinguishable from the standard cosmological model with cosmological constant (i.e. $\Lambda$CDM). In Zhang et al. \cite{zhang}, the Chaplygin inflation has been investigated in the context of Loop Quantum Cosmology, moreover it is shown that the parameters of the Chaplygin inflation model are consistent with the results of 5-year WMAP data.\\
The Modified Chaplygin Gas (MCG) is a generalization of the GCG  with the addition of a barotropic term \cite{jami,jami2,jami3,jami4}. The MCG seems to be consistent with the 5-year WMAP data and henceforth supports the unified model with DE and DM based on Generalized Chaplygin Gas.\\
The MCG is defined as \cite{jami}:
\begin{equation}
p_{\Lambda}=A\rho_{\Lambda}-\frac{D}{\rho_{\Lambda}^\alpha}, \label{mcg1}
\end{equation}
where $A$, $D$ and $\alpha$ are positive constants (with $0 \leq \alpha \leq 1$).\\
The energy density $\rho_D$ of the MCG, calculated using the density conservation equation, is given by:
\begin{equation}
\rho_{\Lambda}=\Big[\frac{D}{1+A}+\frac{B}{a^{3(1+\alpha)(1+A)}}\Big]^{\frac{1}{1+\alpha}},\label{mcg2}
\end{equation}
where $B$ represents a constant of integration. I now want to reconstruct the potential and dynamics of the scalar field $\Phi$ in the light of LECRDE, for this purpose I proceed as,
since I know by string theory that for a homogeneous and time dependent scalar field $\Phi$ energy density and pressure are defined by:
\begin{eqnarray}
\rho_{\Lambda}&=&\frac{\sigma}{2}\dot\Phi^2+V(\Phi),\label{mcg3}\\
p_{\Lambda}&=&\frac{\sigma}{2}\dot\Phi^2-V(\Phi).\label{mcg4}
\end{eqnarray}
The EoS paramater of MCG $\omega_{\Lambda}$ is given by:
\begin{equation}
\omega_{\Lambda}=\frac{p_{\Lambda}}{\rho_{\Lambda}}=\frac{\sigma \dot\Phi^2-2V(\Phi)}
{\sigma \dot\Phi^2+2V(\Phi)}.\label{mcg5}
\end{equation}
Using Eqs. (\ref{mcg3}), (\ref{mcg4}) and (\ref{mcg5}), I get the kinetic energy $\dot{\Phi}^2$ and the scalar potential $V\Big(\Phi\Big)$ terms, respectively, as:
\begin{eqnarray}
\dot{\Phi}^2&=&\frac{1}{\sigma}(1+\omega_{\Lambda})\rho_{\Lambda},\label{mcg6}\\
V\left(\Phi\right)&=&\frac{1}{2}(1-\omega_\Lambda)\rho_{\Lambda}.\label{mcg7}
\end{eqnarray}
Using Eq. (\ref{mcg1}), the EoS parameter $\omega_{\Lambda}$ of MCG can be written as:
\begin{equation}
\omega_{\Lambda}=\frac{p_{\Lambda}}{\rho_{\Lambda}}=A-\frac{D}{\rho_{\Lambda}^{\alpha +1}}.\label{mcg8}
\end{equation}
From Eq. (\ref{mcg8}), I derive that the parameter $D$ is given by:
\begin{equation}
D=\rho_{\Lambda}^{\alpha +1}\Big(A-\omega_{\Lambda}\Big). \label{mcg9}
\end{equation}
Inserting the EoS parameter $\omega_{\Lambda}$ of the LECRDE given in Eq. (\ref{22}) and the EoS parameter $\omega_{\Lambda}$ of the PLECRDE given in Eq. (\ref{rplechde}) into Eq. (\ref{mcg9}), I obtain, respectively:
\begin{eqnarray}
D &=&\Big[\rho_{\Lambda}\Big]^{\alpha +1}\Big [A-3\varepsilon + \frac{M^2_p/3}{3\alpha
M^2_p+\gamma_1R\ln(M^2_p/R)+\gamma_2R} - \frac{1}{3}\Big(\frac{1+\Omega_k}{\Omega_{\Lambda}}\Big)
\Big],\label{mcg10}\\
D &=&\Big[\rho_{\Lambda}\Big]^{\alpha +1}\Big [A-3\varepsilon +\frac{1}{3\Big(3\alpha -\beta R^{\gamma /2-1}\Big)}-\frac{1}{3}\Big(\frac{1+\Omega_k}{\Omega_{\Lambda}}\Big)
\Big].\label{mcg11}
\end{eqnarray}
From Eq. (\ref{mcg2}), I can derive the following expression for the constant of integration $B$:
\begin{equation}
 B=a^{3\left(\alpha +1\right)\left(A+1\right)}\left(\rho _{\Lambda }^{\alpha +1}-\frac{D}{1+A}\right), \label{bBb}
\end{equation}
which can be written, using Eq. (\ref{mcg9}), as:
\begin{eqnarray} B=\left[a^{3\left(A+1\right)}\rho_{\Lambda}\right]^{1+\theta}\left(\frac{1+\omega_{\Lambda}}{1+A}\right). \label{BBB}
\end{eqnarray}
Inserting the EoS parameter $\omega_{\Lambda}$ of the LECRDE given in Eq. (\ref{22})and the EoS parameter $\omega_{\Lambda}$ of the PLECRDE given in Eq. (\ref{rplechde}) in Eq. (\ref{BBB}), I obtain the following expressions for $B$:
\begin{eqnarray}
B &=& \frac{\left[a^{3\left(1+A\right)}\rho_{\Lambda}\right]^{\theta
+1}}{1+A}\left[1+3\varepsilon -
\frac{M^2_p/3}{3\alpha M^2_p+\gamma_1R\ln(M^2_p/R)+\gamma_2R} +
\frac{1}{3}\left(\frac{1+\Omega_k}{\Omega_{\Lambda}}\right)
  \right],\label{mcg12} \\
B &=& \frac{\left[a^{3\left(1+A\right)}\rho_{\Lambda}\right]^{\theta
+1}}{1+A}\left[1+3\varepsilon -\frac{1}{3\left(3\alpha -\beta R^{\gamma /2-1}\right)}+\frac{1}{3}\left(\frac{1+\Omega_k}{\Omega_{\Lambda}}\right)
  \right].\label{mcg13}
\end{eqnarray}

Using Eqs. (\ref{mcg6}), (\ref{mcg7}), (\ref{mcg10}) and (\ref{mcg12}), I obtain the kinetic and potential terms for the LECRDE model as:
\begin{eqnarray}
\sigma \dot\Phi^2
&=&\rho_{\Lambda}\Big[1+3\varepsilon -
\frac{M^2_p/3}{3\alpha M^2_p+\gamma_1R\ln(M^2_p/R)+\gamma_2R} +
\frac{1}{3}\left(\frac{1+\Omega_k}{\Omega_{\Lambda}}\right)
\Big],\label{mcg14} \\
2V(\Phi)&=&\Big[1-3\varepsilon +
\frac{M^2_p/3}{3\alpha M^2_p+\gamma_1R\ln(M^2_p/R)+\gamma_2R} -
\frac{1}{3}\left(\frac{1+\Omega_k}{\Omega_{\Lambda}}\right)\Big]\label{mcg15}
\end{eqnarray}

Moreover, using Eqs. (\ref{mcg6}), (\ref{mcg7}), (\ref{mcg11}) and (\ref{mcg13}), I obtain the kinetic and potential terms for the PLECRDE model as:
\begin{eqnarray}
\sigma \dot\Phi^2 &=&\rho_{\Lambda}\Big[1+3\varepsilon -\frac{1}{3\left(3\alpha -\beta R^{\gamma /2-1}\right)}+\frac{1}{3}\left(\frac{1+\Omega_k}{\Omega_{\Lambda}}\right)  \Big],\label{mcg14-1}\\
2V(\Phi)&=&\Big[1-3\varepsilon+\frac{1}{3\left(3\alpha -\beta R^{\gamma /2-1}\right)}-\frac{1}{3}\left(\frac{1+\Omega_k}{\Omega_{\Lambda}}\right)
\Big] \label{mcg15-1}
\end{eqnarray}

Since $\dot\Phi=\Phi'H$ (where the prime denotes differentiation with respect to the cosmic time $x=\ln a$), I obtain, from Eq. (\ref{mcg14}), that the evolutionary form of the MCG for the LECRDE model is given by:
\begin{eqnarray}
\Phi(a) - \Phi(a_0) =
\int_{a_0}^{a}\Big\{\Big[\frac{3M_p^2\Omega_{\Lambda}}{\sigma}\Big(1+3\varepsilon -
\frac{M^2_p/3}{3\alpha M^2_p+\gamma_1R\ln(M^2_p/R)+\gamma_2R} +
\frac{1}{3}\left(\frac{1+\Omega_k}{\Omega_{\Lambda}}\right)
\Big)\Big]\Big\}^{1/2} \frac{da}{a}. \label{mcg16}
\end{eqnarray}

Moreover, using Eq. (\ref{mcg14-1}), I have that the evolutionary form of the MCG for the PLECRDE model is given by:
\begin{eqnarray}
\Phi(a) - \Phi(a_0) =   \int_{a_0}^{a}\Big\{\Big[\frac{3M_p^2\Omega_{\Lambda}}{\sigma}\Big(1+3\varepsilon -\frac{1}{3\left(3\alpha -\beta R^{\gamma /2-1}\right)}+\frac{1}{3}\left(\frac{1+\Omega_k}{\Omega_{\Lambda}}\right) \Big)\Big]\Big\}^{1/2} \frac{da}{a}. \label{mcg17}
\end{eqnarray}

Eqs. (\ref{mcg16}) and (\ref{mcg17}) represent the reconstructed potentials of the MCG model.\\
In the limiting case for flat dark dominated universe, i.e. when $\gamma_1=\gamma_2=0$, $\Omega_{\Lambda}=1$ and $\Omega_k$=0 for the LECRDE model and $\beta=0$, $\Omega_{\Lambda}=1$ and $\Omega_k$=0 for the PLECRDE model, the scalar field and the potential of the GCG reduce to, respectively:
\begin{eqnarray}
    \Phi(t) &=& \frac{6\alpha M_p}{\Big(12\alpha -1  \Big)}\sqrt{\frac{12\alpha -1  + 27\alpha \varepsilon}{3\alpha \sigma}}\ln \Big( t \Big), \label{mcg20}\\
V(\Phi) &=& \frac{54\alpha^2 M_p^2}{\Big( 12\alpha - 1 \Big)^2} \Big( \frac{6\alpha  - 27\alpha \varepsilon +1}{9\alpha} \Big)\frac{1}{t^2}. \label{mcg21}
\end{eqnarray}
In the limiting case of $\varepsilon = 0$, Eqs. (\ref{mcg20}) and (\ref{mcg21}) become, respectively:
\begin{eqnarray}
    \Phi\left(t\right) &=& \frac{6\alpha M_p}{\sqrt{3\alpha \sigma \Big( 12\alpha -1\Big)}}\ln \Big( t \Big), \label{mcg22}\\
V\left(\Phi\right)&=&\frac{6\alpha(6\alpha+1)}{(12\alpha-1)^2}M_p^2\exp{\Big[\frac{-\sqrt{3\alpha(12\alpha-1)}}{3\alpha
M_p}\Phi\Big]}.\label{mcg23}
\end{eqnarray}



\section{CONCLUSIONS}

In this paper, I studied the entropy-corrected and the power law entropy corrected versions of the HDE model (which is an attempt to prove the nature of DE within the framework of Quantum Gravity). I considered the DE in interaction with DM in the non-flat FRW universe and I choose as IR cut-off the Ricci scalar $R$. Moreover, I also inserted a logarithmic correction term to the energy density $\rho_{\Lambda}$ of entropy corrected HDE model and I also considered the presence of dissipative effects due to the presence of bulk viscosity in cosmic
fluids. The addition of the correction term is motivated from the Loop Quantum Gravity (LQG), which is one of the most promising theories of Quantum Gravity. Using the energy densities of the two models considered, I obtained the EoS parameter $\omega$, the deceleration parameter $q$ and evolution of energy density parameter $\Omega ' _{\Lambda}$ for the interacting LECRDE and PLECRDE models. Moreover, I established a correspondence between the two models and the Modified Chaplygin Gas (MCG), the tachyon, K-essence, dilaton and quintessence field models in the hypothesis of non-flat FRW universe. These correspondences are important because they allow us to understand how the various candidates of DE are mutually related to each other. The limiting case for the flat dark dominated universe (i.e. when $\gamma_1=\gamma_2=0$, $\Omega_{\Lambda}=1$ and $\Omega_k$=0 for the LECRDE model and $\beta=0$, $\Omega_{\Lambda}=1$ and $\Omega_k=0$ for the PLECRDE model) were studied in each scalar field and I found that the EoS parameter $\omega$ is constant in this case. I also calculated the scalar field and its potential, which can be obtained by idea of power-law expansion of scalar field. The results obtained are also in agreement with other works previously done in the same field. \\

\end{document}